\documentclass[twocolumn,aps,prb]{revtex4} 
\usepackage{graphics,amsmath,graphicx}  

\begin{document}
\title{Phenomenological model for charge dynamics and optical response of disordered systems: 
application to organic semiconductors.}

\author{S. Fratini$^{1,2}$, S. Ciuchi$^3$, D. Mayou$^{1,2}$} 

\affiliation{$^1$Univ. Grenoble Alpes, Inst. NEEL, F-38042 Grenoble, France \\
$^2$CNRS, Inst NEEL, F-38042 Grenoble, France\\
$^3$Dipartimento di Scienze Fisiche e Chimiche\\ 
Universit\`a dell'Aquila, CNISM and Istituto Sistemi Complessi CNR, 
via Vetoio, I-67010 Coppito-L'Aquila, Italy}

\begin{abstract}
We provide a phenomenological formula which describes the low-frequency optical absorption of
charge carriers in disordered systems with localization. This 
allows to extract, from experimental data on the optical 
conductivity, the relevant microscopic parameters 
determining the transport properties, such as
the carrier localization length and  the elastic  and inelastic scattering times.
This general formula is  tested and 
applied here to organic semiconductors, where dynamical molecular disorder 
is known to play a key role in the transport properties.
The present treatment captures the basic ideas underlying 
the recently proposed transient localization scenario for charge transport,  
extending it from the d.c. mobility to the frequency domain. 
When applied to existing optical measurements in rubrene FETs, our analysis provides quantitative 
evidence for the transient localization phenomenon.  
Possible applications to other disordered electronic systems are briefly discussed.
\end{abstract}

\date{\today}
\maketitle
\section{Introduction}

The semiclassical Bloch-Boltzmann transport theory, which
relies on the existence of well-defined extended "band" states that are weakly scattered
by impurities and phonons, is known to provide a successful description of 
charge carrier  dynamics 
in ordinary wide-band semiconductors.
It has been shown in recent years \cite{Troisi06,Zuppi,Ciuchi11} that such standard paradigm 
is not appropriate to organic semiconductors, which  are 
more effectively described by taking the strong disorder limit as a starting point.
The reason for this is that even in ultrapure crystalline
samples where extrinsic sources of disorder are removed, 
 large thermal molecular
motions arise due to the weak van der Waals intermolecular binding. Such dynamical 
fluctuations in molecular positions and orientations strongly scatter the charge carriers, 
causing a breakdown of the assumptions 
underlying semiclassical transport.\cite{Friedman,Cheng,Fratini09,Ciuchi11} 
Unlike static disorder, however,  dynamical disorder is unable to fully localize the carriers: 
after a {\it transient localization} regime which extends up to the typical timescale of molecular vibrations,
a diffusive behavior is eventually established. 
 It can be shown that the  resulting d.c. mobility 
 is a decreasing 
 function of temperature, with a  power-law behavior which 
resembles that of semiclassical "band-like" carriers.
Its modest value, however,
at best of the order of a few tens of $ cm^2/Vs$ at room temperature,
is there to remind us of the presence of an underlying strong disorder.

A more direct signature of 
this unconventional transport mechanism
is predicted in the a.c. response of the carriers: associated to the transient localization phenomenon,
a peak emerges in the optical conductivity at a frequency related to
the amount of molecular disorder, 
deeply modifying the usual Drude response expected for 
band-like carriers.\cite{Ciuchi11,Cataudella}
All these features --- power-law temperature dependence, low values of the mobility and the existence 
of a localization peak in the optical conductivity ---  are commonly found in experiments  
on high-mobility organic semiconductors, giving support to the transient localization scenario for charge
transport.

It is our aim here to derive a general phenomenological formula describing the low-frequency optical absorption of
charge carriers in disordered systems with localization. Such a formula should be able to 
provide a theoretically simple  description 
of the transient localization scenario for organic semiconductors,  capturing the main features evidenced in recent 
numerical simulation studies. 
Rather than studying
a particular microscopic model, 
as was done in Refs. \cite{Troisi06,Zuppi,Fratini09,Ciuchi11,Cataudella,Ciuchi12},
we therefore introduce a phenomenological model for the carrier dynamics which
yields a closed analytical form for both the d.c. mobility and optical conductivity. Our model 
is able to interpolate
between the Drude-like response of diffusive carriers and the absorption peak
expected in the presence 
of strong Anderson localization, and establishes a direct connection between the temperature dependent 
mobility and the optical conductivity. 
We then perform a benchmarking of the phenomenological formula for the optical conductivity by comparing it with  
a true microscopic calculation of this quantity in a model system \cite{Ciuchi12}. Our analysis demonstrates 
that the formula derived here can be used  to accurately extract the
microscopic parameters governing the carrier dynamics,  such as 
the transient localization length and the relevant scattering rates,
and to estimate the electron-intermolecular vibration coupling strength and the amount of extrinsic (i.e. non-thermal) 
disorder. This allows us to analyze  quantitatively the  optical absorption spectra available in organic FETs
in the framework of the transient localization scenario.
In the conclusive section we briefly discuss how the present theory can be applied not only to organic semiconductors but also  
to bad metals and other disordered systems, including organic conductors and carbon nanotubes.  

For readers not interested in the formal developments of the theory, the central formula of the paper 
which can be used to fit optical conductivity data in disordered semiconductors is 
presented in Sec. II D. An analogous formula valid for degenerate electron systems at low temperatures is given 
in Appendix C.

\section{Phenomenological model for the charge dynamics}

\subsection{General formalism}
We start by briefly reviewing a recently developed theoretical 
framework \cite{Mayou88,Khanna,Roche,Mayou00,Triozon,Trambly06,Ciuchi11,Ciuchi12}  
based on the Kubo formula, that relates the quantum diffusion of electrons and the optical conductivity. 
This formalism has been successfully applied to analyze the carrier dynamics in 
electronic systems where localization effects cause a breakdown of usual Boltzmann transport
including quasicrystals\cite{Mayou00,Trambly06}, organic semiconductors\cite{Ciuchi11,Ciuchi12},
and graphene \cite{Trambly13}. 
% In what follows we shall use the notations of Ref. \cite{Ciuchi12}.   

The key ingredient of such formalism is the quantum-mechanical spread 
$\Delta X^2(t)=\langle \lbrack \hat{X}(t)-\hat{X}(0)\rbrack^2\rangle $ of the position operator  
$\hat{X}(t)=\sum_{i=1}^N \hat{x}_i(t)$ of an N-electron system, which  contains all  
the information on the electron dynamics over time.
In particular, the first and second derivatives of the electronic spread yield respectively the 
instantaneous diffusivity, 
\begin{equation}
\mathcal{D}(t)=\frac{1}{2}\frac{d\Delta X^2}{dt}= \frac{1}{2}\int_0^tC_+(t^\prime) dt^\prime
\end{equation} 
and the retarded {\it anticommutator} velocity correlation function,  
\begin{equation}
C_+(t)= \frac{d^2\Delta X^2}{dt^2} =\theta(t) \langle \lbrace \hat{V}(t),\hat{V}(0)\rbrace \rangle,
\end{equation} 
with the initial condition $\Delta X^2(t=0)=0$.
Following Ref.\cite{Ciuchi12}, once that the time-dependent quantum-mechanical spread or equivalently the anticommutator
velocity correlation function are known, the usual commutator correlation function that enters the 
Kubo response theory is obtained by imposing the detailed balance condition.\cite{Ciuchi12}
The real part of the optical conductivity is then obtained as 
\begin{equation}
\sigma(\omega)=\frac{e^2\tanh(\frac{\hbar\omega}{2k_BT})}{\hbar \omega \Omega} Re C_+(\omega) 
\label{eq:sigmatransf}
\end{equation}
where $e$ is the electron charge, $\Omega$ is the system volume and 
 $C_+(\omega)=\int_0^\infty e^{i\omega t} C_+(t) dt$. A similar formula has been proposed in Ref. \cite{Lindner10} to account for 
the bad metallic behavior in  a system of hard-core bosons.

In the following sections we shall focus on the non-degenerate low-density limit appropriate to 
weakly doped semiconductors. In this case the correlation function $C_+(t)$ as well as the quantum spread 
are directly proportional to the number of carriers $N$, being thermodynamical averages for $N$ independent  particles. 
We present a phenomenological ansatz for the correlation function $C_+(t)$ and the  quantum diffusion of electrons in 
organic semiconductors and derive the corresponding optical conductivity lineshape.
The modifications of the formalism which apply to degenerate electron systems are presented in Appendix C.

\subsection{Localized carriers}

\begin{figure}[h]
   \centering
\includegraphics[width=8.cm]{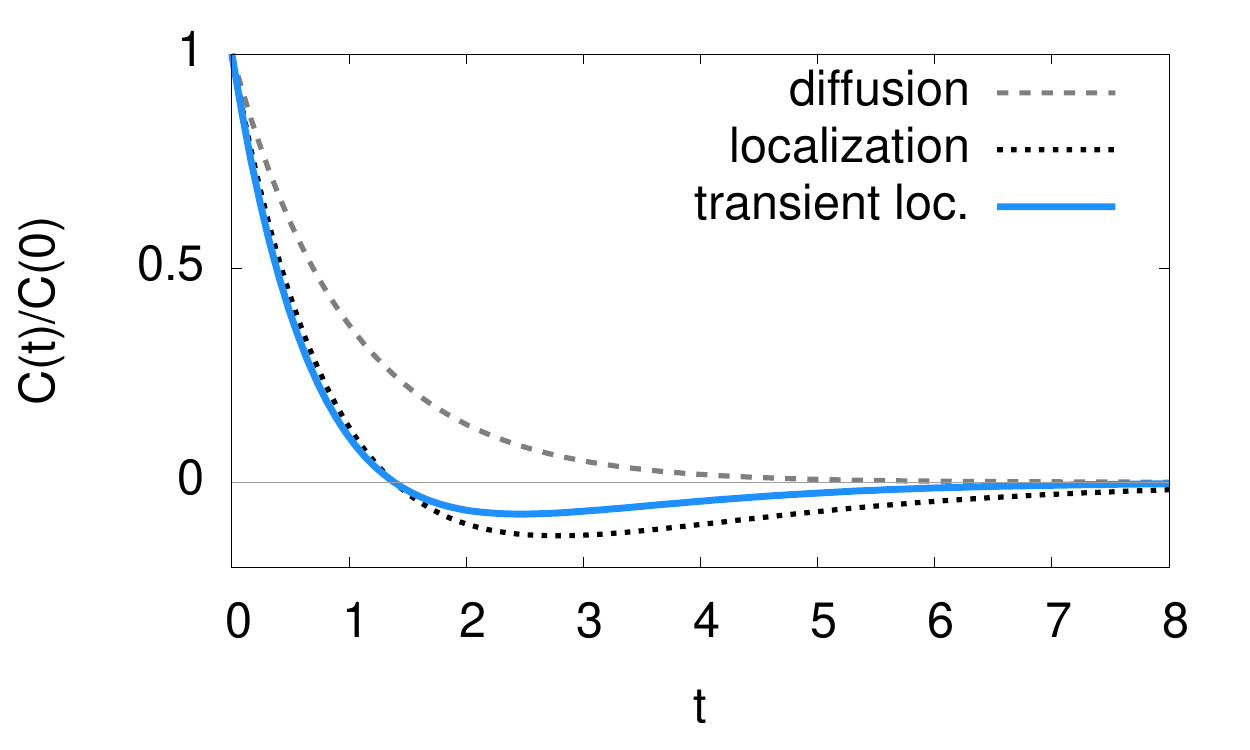}
\includegraphics[width=8.cm]{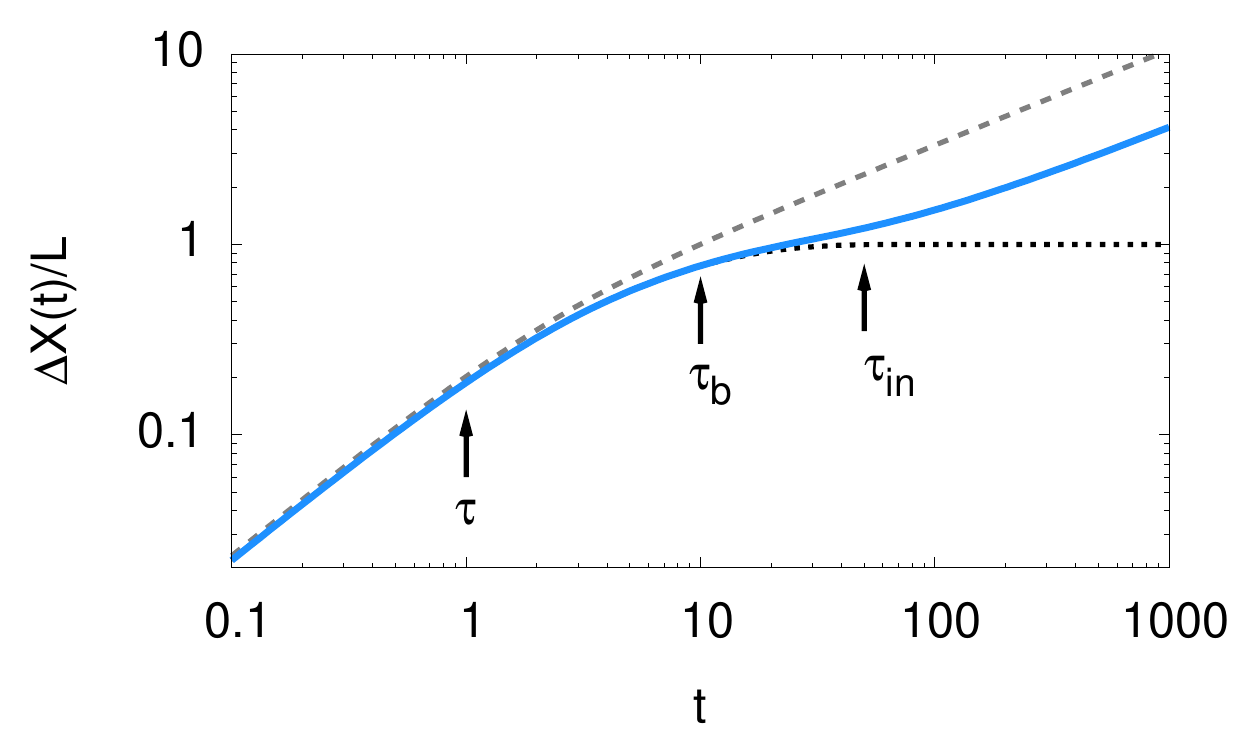}
   \caption{(a) The velocity correlation function 
   $C(t)$  obtained from the phenomenological RTA for  full localization  
   [Eq.(\ref{eq:Cloc}), black dotted line] and transient localization [Eq.(\ref{eq:Cdyn}), full blue line]. 
   We have taken $\tau=1$, $\tau_b=2$ and  $\tau_{in}=5$ (times are in units of $\tau$).  
   The diffusive term alone with $\tau=1$ is shown for comparison [first term in Eq.(\ref{eq:Cloc}),
   dashed gray line];
   (b) The corresponding quantum spread $\Delta X(t)$ per particle, in units of the localization length $L$. 
   For illustrative purposes, we have taken here 
   $\tau=1$, $\tau_b=10$ and  $\tau_{in}=50$
    (note the logarithmic scale on the time axis). Arrows indicate the three characteristic timescales 
    (elastic scattering, backscattering and inelastic). }
 \label{fig:XandC}
 \end{figure}

Our starting point is the following reference model, that accounts  for  carrier localization
in the limit of strong static disorder (from now on we shall always refer
 to the  anticommutator velocity correlation function and drop the subscript $+$ for simplicity): 
\begin{eqnarray}
& & C(t)= \frac{C(0)}{1/\tau-1/\tau_b}\left\lbrack 
\frac{1}{\tau}e^{-t/\tau}-\frac{1}{\tau_b}e^{-t/\tau_b}\right\rbrack 
\label{eq:Cloc}\\
& & \Delta X^2(t)=  \frac{C(0)}{1/\tau-1/\tau_b}\left\lbrack 
\tau_b(1-e^{-t/\tau_b})-\tau(1-e^{-t/\tau})
\right\rbrack 
\label{eq:X2loc}.
\end{eqnarray}
The correlation function in  Eq. (\ref{eq:Cloc}) consists of two terms. A first exponential decay
causes relaxation of the velocity on a timescale given by the elastic scattering time $\tau$. This is
equivalent to the usual decay term which is present in the semiclassical Boltzmann theory\cite{Gruner},
and which is responsible for the Drude response of the carriers  (see below and  Appendix A). 
A second "backscattering" term, with a timescale $\tau_b>\tau$, 
is introduced in order to describe the negative velocity correlations which 
lead to electron localization at long times. The choice of the prefactors of the 
exponential terms between brackets ensures that the diffusivity vanishes at long times,
$\int_0^\infty C_+(t^\prime) dt^\prime=2\mathcal{D}(t\to \infty)=0$.
This function is illustrated in Fig. \ref{fig:XandC}-a (dotted line).
\cite{note-geometrical}

The  expression Eq. (\ref{eq:X2loc}) of the quantum diffusion  follows from double integration 
of  Eq. (\ref{eq:Cloc}). It describes  three different regimes expected  in a localized
N-electron system in different time ranges, as illustrated in Fig. \ref{fig:XandC}-b (dotted line): 
a ballistic evolution, $\Delta X^2(t)= C(0)t^2/2$,  at initial times is followed by  
diffusion, $\Delta X^2(t) \propto t$, 
setting in after the elastic scattering time, $t>\tau$. 
The diffusive behavior 
is eventually destroyed by backscattering, causing electronic localization, $\Delta X^2(t) \to \mathrm{const}$, 
at  $t> \tau_b$.

From that several relations can be derived whose physical content is particularly instructive.
First of all, from Eq. (\ref{eq:X2loc}), the value of the localization length, 
$L^2=\lim_{t \to \infty}\Delta X^2(t)/ N$, 
fixes the diffusive behavior of the  carriers  prior to localization.
By expanding Eq. (\ref{eq:X2loc}) in the range $\tau<t<\tau_b$ we obtain $\Delta X^2(t)= 2ND_{sc}(t-\tau)$, with 
$D_{sc}=L^2/(\tau_b-\tau)$ the semiclassical diffusivity.\cite{C0}
Observing that generally $\tau_b\gg \tau$ we can rewrite this as $L\simeq\sqrt{D_{sc} \tau_b}$, which
can be actually taken as a definition of the backscattering time: it is the time it takes
to the electron spread to attain the localization length $L$ upon diffusing with a rate $D_{sc}$. 
Similarly, the elastic mean free path is $\ell=\sqrt{D_{sc} \tau}$, so that
\begin{equation}
\frac{L}{\ell}\simeq\sqrt{\frac{\tau_b}{\tau}}.
\label{eq:tauratio}
\end{equation}
The ratio of the localization length to the elastic mean free path 
is actually known from the Thouless relation\cite{Beenakker}: apart from numerical factors, it is equal 
to the number of conduction channels available 
in the system. For one-dimensional conduction with one orbital per unit cell, 
for example, the number of channels is one so that $L/\ell$ and  $\tau_b/\tau$ are both constant and independent
of the disorder strength (cf. Sec. III below).  For anisotropic two-dimensional systems, one can roughly 
apply an analogous argument by defining
an effective width $L_{\perp}$ , which corresponds to the localization length along the transverse  direction. 
The effective number of modes in this case is therefore of the order of $L_{\perp}/a_\perp$, and so is the ratio 
$\sqrt{\tau_b/\tau}$.

A second useful relation follows from the observation that, for independent 
non-degenerate particles of mass $m^*$, the semiclassical diffusivity along a given direction 
is known and given by $D_{sc}=2k_BT\tau/m^*$ (see Appendix A). Equating this to the expression given before Eq. 
(\ref{eq:tauratio}) yields 
\begin{equation}
L^2=\frac{2k_BT }{m^*}\tau(\tau_b-\tau).
\label{eq:L2mstar}
\end{equation}
This relation shows that the localization length $L$ and the backscattering time $\tau_b$, i.e. the two parameters 
governing the localization process, are not independent. The validity of both Eq. (\ref{eq:tauratio}) and Eq. 
(\ref{eq:L2mstar}) will be verified numerically in Sec. III, and their practical relevance when fitting experimental 
data will be demonstrated in Sec IV.

\subsection{Transient localization and d.c. mobility within the RTA}

In organic semiconductors, the dynamical nature of molecular disorder  prevents 
localization of the carriers at long times. A recovery of carrier diffusion is  expected
in the case of dynamical disorder because the time  fluctuations of the disorder potential destroy  
the quantum interferences responsible for carrier localization. 
This effect, which occurs on the typical timescale of molecular motions, that we denote  $\tau_{in}$, 
can be included in our phenomenological model in the spirit of the RTA by setting:
\begin{equation}
C_{RTA}(t)=C(t)e^{-t/\tau_{in}} \label{eq:Cdyn}.
\end{equation} 
We assume that the inelastic timescale is the longest timescale in the problem, 
$\tau_{in}>\tau_b>\tau$.  This is a reasonable assumption
in organic semiconductors because the molecular motions that are 
at the origin of disorder are slow, owing
to the large molecular mass and the weak inter-molecular (van der Waals) restoring forces.
The consequence of this assumption is that the behavior of the 
system at the timescales relevant for the buildup of localization 
is very similar to that of the reference case with static disorder, Eq. (\ref{eq:Cloc}).
The main effect of the extra exponential decay in Eq. (\ref{eq:Cdyn}) is to suppress  the 
long-time backscattering [the second term in Eq. (\ref{eq:Cloc})], 
restoring carrier diffusion at long times \cite{note-geometrical}.
This is best seen in the quantum spread $\Delta X^2(t)$, which is readily obtained 
by integrating Eq. (\ref{eq:Cdyn}) twice over time
(see Appendix B for the full expression). 
As shown in Fig. \ref{fig:XandC}-b (full line), the inclusion of the disorder dynamics 
causes a departure from the behavior of
the reference localized system on a scale $t \sim \tau_{in}$. 
This form qualitatively agrees with the quantum spread obtained via 
quantum-classical simulations of a model with dynamical inter-molecular
disorder (see e.g. Fig.1  in Ref.\cite{Ciuchi11}).

Importantly,  when velocity correlations are allowed to relax via Eq. (\ref{eq:Cdyn})
a diffusive behavior is recovered at long times, with a diffusion constant given by:
\begin{equation}
D\simeq \frac{L^2}{2\tau_{in}} \label{eq:Ddyn}.
\end{equation} 
where we have assumed $\tau_{in}\gg \tau_b,\tau$ (see again Appendix B for a more general expression).
The  form Eq. (\ref{eq:Ddyn}) has a clear physical meaning: it corresponds
to the diffusion of particles hopping on a lengthscale $L$ with a jump rate $1/\tau_{in}$.
%It is analogous to the Thouless diffusivity of Anderson insulators. 
The corresponding mobility is obtained from Einstein's relation as 
\cite{Zuppi,Ciuchi11,Ciuchi12}
\begin{equation}
\mu(T)\simeq \frac{e}{k_BT}\frac{L^2(T)}{ 2\tau_{in}}.
 \label{eq:mu}
\end{equation}

It has been shown in Refs. \cite{Ciuchi11,Ciuchi12} that, if we consistently associate $\tau_{in}$ with 
the typical frequency of the relevant inter-molecular vibrations,  $1/\tau_{in}\simeq \omega_0$,
Eq. (\ref{eq:mu}) correctly describes both the absolute value and temperature 
dependence of the mobility observed in crystalline organic semiconductors in the intrinsic regime.
In this case $L$ is a smoothly decreasing function of $T$ because the main source of disorder is 
constituted by molecular motions of thermal origin, and
the localization length decreases upon increasing the amount of disorder\cite{Zuppi,Fratini09,Ciuchi12}.
This leads to an overall power-law temperature dependence of the mobility, which is 
a common feature observed in pure samples at sufficiently high temperatures.

The power-law temperature dependence described above should not be confused with standard semi-classical transport:  
the transient localization form Eq. (\ref{eq:mu}) of the mobility is very different
from the usual semiclassical "Drude" form, $\mu(T)=\frac{e\tau(T)}{m}$, because it arises from 
a fundamentally different microscopic mechanism. 
As can be seen in Fig. \ref{fig:XandC}-b, the diffusivity in the presence of transient localization 
(full line) is generally lower than that of semiclassical carriers (dashed line; see also the $\omega\to 0$ limit
in Fig. 2). 
It is then easy to understand that Eq. (\ref{eq:mu}) can 
describe mobilities that go below the so-called Mott-Ioffe-Regel limit $\mu_0\approx e a^2/\hbar$, which 
is where  the apparent mean-free path 
falls below the typical inter-molecular distance, implying a breakdown of the
semiclassical approximation.
Organic semiconductors are a particularly favorable ground for this breakdown to occur,
because of the large thermal molecular disorder (leading to a short $L$) together with 
 large values of the molecular mass (implying a large $\tau_{in}$), 
both contributing to reduce the value of $\mu$ in Eq. (\ref{eq:mu}). Indeed,  
observing that $L$ reduces to few lattice spacings  at room temperature even in pure samples 
(see Refs. \cite{Fratini09,Ciuchi12} and Fig. \ref{fig:fits} below), 
a sufficient condition for the breakdown of the semi-classical limit is that $\hbar /\tau_{in}< k_B T $,
which is easily reached in these compounds.
The quantitative microscopic calculations of Ref. \cite{Ciuchi12} (see Fig. 2-a there, where the mobility
is conveniently expressed in units of $\mu_0$) do confirm that the Mott-Ioffe-Regel limit is attained 
in pure samples around room temperature, and that the mobility always lies below this
 limit when  sizable
extrinsic disorder is present.

\subsection{Drude-Anderson model for the optical conductivity}

 \begin{figure}[h]
 \includegraphics[width=10.cm]{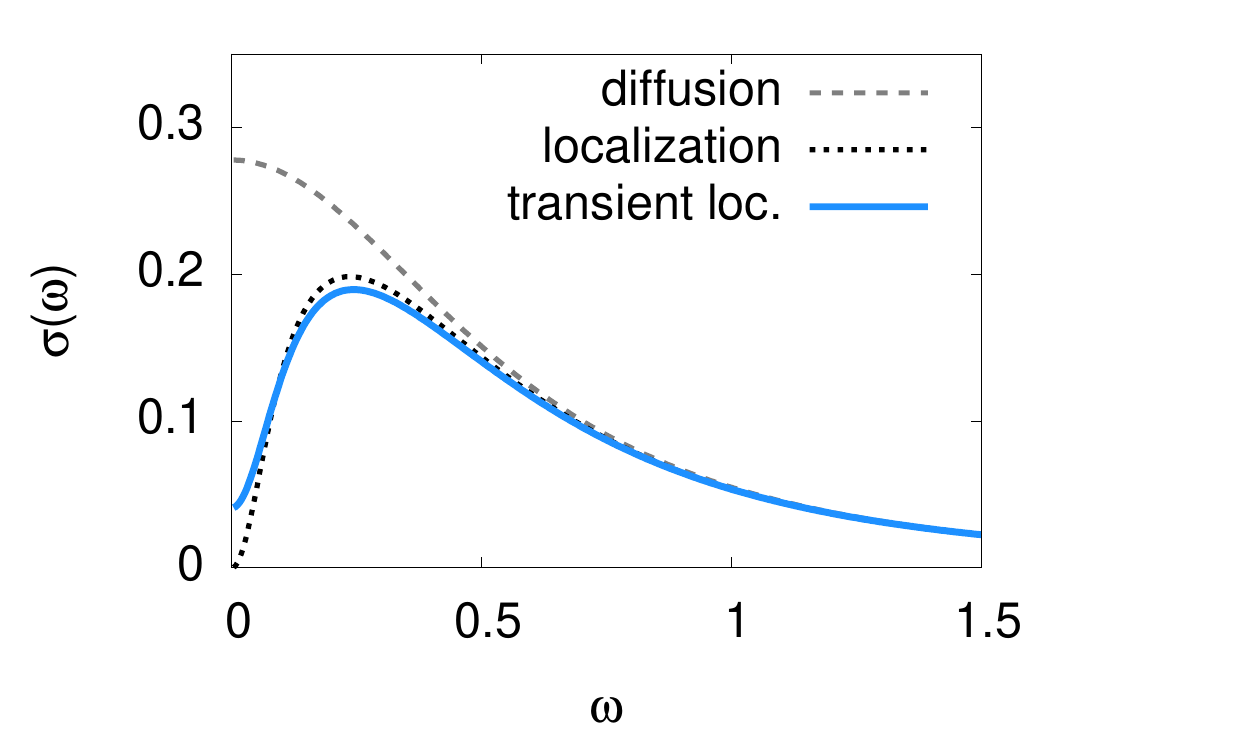}
   \caption{The real  part of the optical conductivity  Eq. (\ref{eq:optcond}):  
   full line, $\tau=1$, $\tau_b=10$, $\tau_{in}=50$ and $k_BT=0.2\hbar/\tau$ (frequencies in units of $1/\tau$). 
   As in Fig. 1,
   the black dotted line is the localization limit, obtained for $\tau_{in}\to\infty$, and the gray dashed line is the 
   diffusive response alone. 
   }
 \label{fig:sigmaphenom}
 \end{figure}
We are now in a position to express the optical conductivity corresponding to 
the phenomenological model Eq. (\ref{eq:Cdyn}).
 From Eq. (\ref{eq:sigmatransf}) 
we can write
\begin{eqnarray}
\sigma(\omega)&=&\frac{ne^2L^2}{\tau_b-\tau}\frac{\tanh(\frac{\hbar\omega}{2k_BT})}{\hbar\omega} \times 
\label{eq:optcond}\\
& & \times Re \ \left\lbrack
\frac{1}{1+\tau/\tau_{in}-i\omega\tau}-\frac{1}{1+\tau_b/\tau_{in}-i\omega\tau_b}
\right\rbrack \nonumber 
\end{eqnarray}
with $n=N/\Omega$.
The above expression ensures that $ \sigma(\omega)\ge 0$ at all frequencies. 
This can be easily shown in the static case $\tau_{in}\to \infty$ 
by taking explicitly the real part in Eq. (\ref{eq:optcond}). The extension of the proof 
to the dynamic case follows by observing that the  product in 
Eq. (\ref{eq:Cdyn}) implies a Lorentzian convolution in frequency space, so that $\sigma(\omega)$ remains 
positive-definite.

As is illustrated in Fig. \ref{fig:sigmaphenom}  (see also Appendix A),
the lineshape described by Eq. (\ref{eq:optcond}) 
actually interpolates 
between the Drude-like response of diffusive carriers and the 
finite-frequency peak expected in the presence of Anderson localization
--- we therefore call it Drude-Anderson formula. 
The shape of $\sigma$ in Fig. \ref{fig:sigmaphenom}-a can be easily understood following the
discussion of the velocity correlation function after Eqs. (\ref{eq:Cloc}) and  (\ref{eq:Cdyn}).
Starting from a typical Lorentzian diffusive response of  
width $\sim 1/\tau$ [the first term between brackets in Eq. (\ref{eq:optcond}), shown
as a dashed line], 
the backscattering correction (the second term between brackets) 
causes a suppression of spectral weight at low frequencies, on a scale determined by  
$1/\tau_b$. The usual monotonic Drude-like response obtained for semiclassical transport
is therefore transformed into a characteristic {\it localization peak}\cite{MottKaveh,Smith}, 
whose position is ruled by the backscattering rate $1/\tau_b$, and whose 
high frequency tails are controlled by the elastic scattering rate $1/\tau$. 
In the case of static disorder ($\tau_{in}\to \infty$), 
the suppression of conductivity is complete at zero frequency, where
carrier localization implies $\sigma(0) = 0$. Disorder dynamics 
%provide an additional frequency scale to the problem. 
restore a finite
d.c. conductivity, which is achieved via a transfer of spectral weight from the localization peak 
to the narrow window $0\le \omega\lesssim 1/\tau_{in}$. 
In this frequency interval, 
the optical conductivity saturates to the d.c. value $\sigma_{d.c.}(T)\simeq (ne^2/2k_BT) L^2 /\tau_{in}$, 
as can be checked by taking 
the limit $\omega\to 0$ in Eq. (\ref{eq:optcond}).
This of course agrees with Eq. (\ref{eq:mu}), as can be checked by applying the low-density
expression\cite{Ciuchi12} $\mu=\sigma_{d.c.}/(ne)$.

A simpler expression for the optical absorption can be derived in the relevant case where the 
three timescales are well separated, i.e. when $\tau\ll \tau_b\ll\tau_{in}$. In this case, in 
the frequency interval 
$1/\tau_{in} \lesssim \omega \lesssim 1/\tau$ around the peak region we can write 
\begin{eqnarray}
\label{eq:optcondsimp}
\sigma(\omega)&\simeq &\frac{ne^2L^2}{\tau_b-\tau}\frac{\tanh(\frac{\hbar\omega}{2k_BT})}{\hbar\omega} 
\frac{(\omega \tau_b)^2}{1+(\omega \tau_b)^2}. 
\end{eqnarray}
The corresponding lineshape now only depends on two parameters, the backscattering time $\tau_b$ and the 
temperature $T$. The following  expressions for the peak position $\omega^*$ can be obtained
in the two regimes of low and high temperatures compared to the backscattering rate:
\begin{eqnarray}
\omega^*& =&1/\tau_b \; \; \; \;\;\;\; \; \; \;\;\;\; k_BT \lesssim 0.3 \hbar/\tau_b
 \label{eq:wstar1}\\
\omega^* &=&  12^{1/4} \sqrt{k_BT/\hbar \tau_b} \; \; \;  k_BT \gtrsim 0.3 \hbar/\tau_b.
 \label{eq:wstar2}
\end{eqnarray}
%12^{1/4}=1.86
Eq. (\ref{eq:wstar2}) applies to
the intrinsic transport regime of organic semiconductors, as shown below. This expression
can be  useful in practice, as it provides a rapid rule to 
estimate the backscattering rate directly from the 
position of the peak in the optical conductivity.

\section{Theoretical benchmarking}

 \begin{figure}[h!]
   \centering
\includegraphics[width=6.cm]{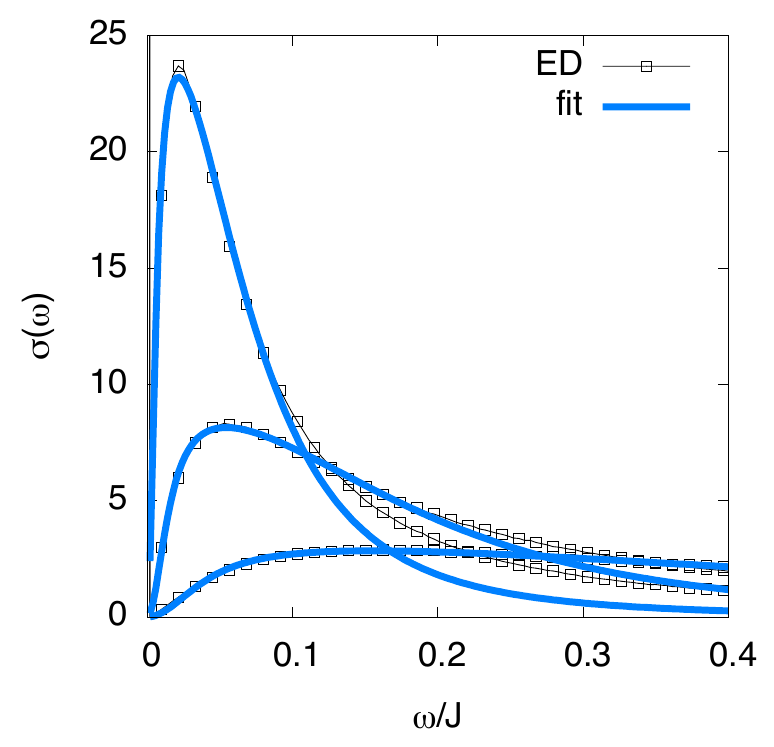}
\includegraphics[width=6.cm]{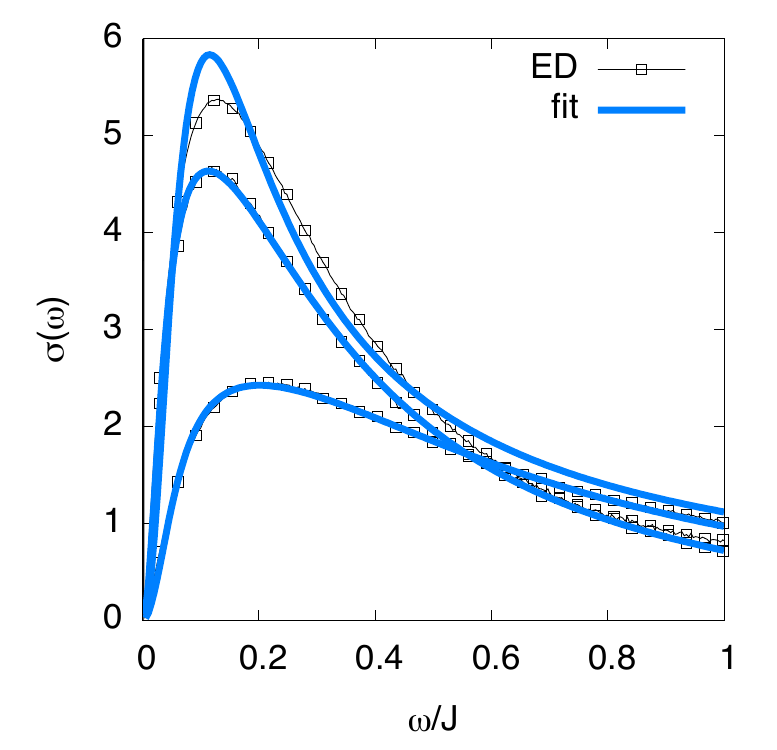}
   \caption{(a) Optical conductivity per particle 
   calculated by ED on a one-dimensional clean chain (no extrinsic site
   disorder) with thermal inter-molecular disorder
   in the static limit $1/\tau_{in}=0$ (gray squares). The full blue lines are 
   the corresponding fits with Eq. (\ref{eq:optcond}). 
   Parameters are $\lambda=0.17$ and $T=0.05;0.1;0.2 J$ (from highest to lowest peak); 
   the unit of conductivity is the Mott-Ioffe-Regel value $\sigma_0=ne^2a^2/\hbar$, with $a$ the 
   inter-molecular distance and $n$ the density (see text); 
    frequencies are expressed in units of $J/\hbar$, the inter-molecular transfer rate;   
    (b) Same, with extrinsic site disorder
   $\Delta=0.2 J$ included.}
 \label{fig:sigma}
 \end{figure}

In order to provide a benchmark for its practical use in the analysis of experiments,  
here we test our formula on the results of exact diagonalization (ED) studies of a 
model system that has been successfully applied\cite{Troisi06,Fratini09,Ciuchi11,Ciuchi12} to 
address the microscopic transport mechanism in organic semiconductors.
By performing fits of the exactly calculated spectra we are able to check that formula Eq. (\ref{eq:optcond}) 
allows to consistently extract the microscopic parameters of the theory, and that it 
accurately recovers those calculated independently within the model when these are known. 
It can therefore be 
used with confidence as a simple and powerful tool for the analysis of experiments.

The model\cite{Troisi06,Fratini09,Ciuchi11,Ciuchi12} 
\begin{eqnarray}
H=\sum_i \epsilon_i c^+_ic_i - \sum_{\langle ij \rangle} J_{ij} (c^+_ic_j + H.c.)
\end{eqnarray}
considers one-dimensional conduction in 
the presence of 
disorder in the inter-molecular transfer integrals $J_{ij}$ (of average $J$), arising  
due to inter-molecular displacements of thermal origin. 
Such intrinsic disorder is governed by the temperature $T$, 
that sets the amplitude of inter-molecular vibrations, 
and by the dimensionless coupling $\lambda$, that controls how inter-molecular motions affect the electronic 
states (see e.g. Ref.\cite{Ciuchi12}). Site disorder is also included, 
representing extrinsic electrical potentials $\epsilon_i$ originating from impurities and defects.\cite{Ciuchi12,Minder14}. 
 The amount of extrinsic disorder is characterized by the variance $\Delta$ of the site energy distribution,
 that is taken to be Gaussian. 
 The model is solved in the static limit, corresponding to
$1/\tau_{in}=0$, where both the $\epsilon_i$ and $J_{ij}$ are time-independent variables. 
As shown in Fig. 2-a this assumption 
does not change significantly the lineshape in the peak region, and it has the advantage of allowing for 
an exact solution of the problem, which is done here 
via exact diagonalization on a 256-site chain and subsequent 
averaging over a large number 
of disorder configurations (up to $100000$). 
Full details on the microscopic model and the method of solution can be found in Ref.
\cite{Ciuchi12}.

\begin{widetext}
 \begin{figure*}[ht]
   \centering
\includegraphics[width=5.8cm]{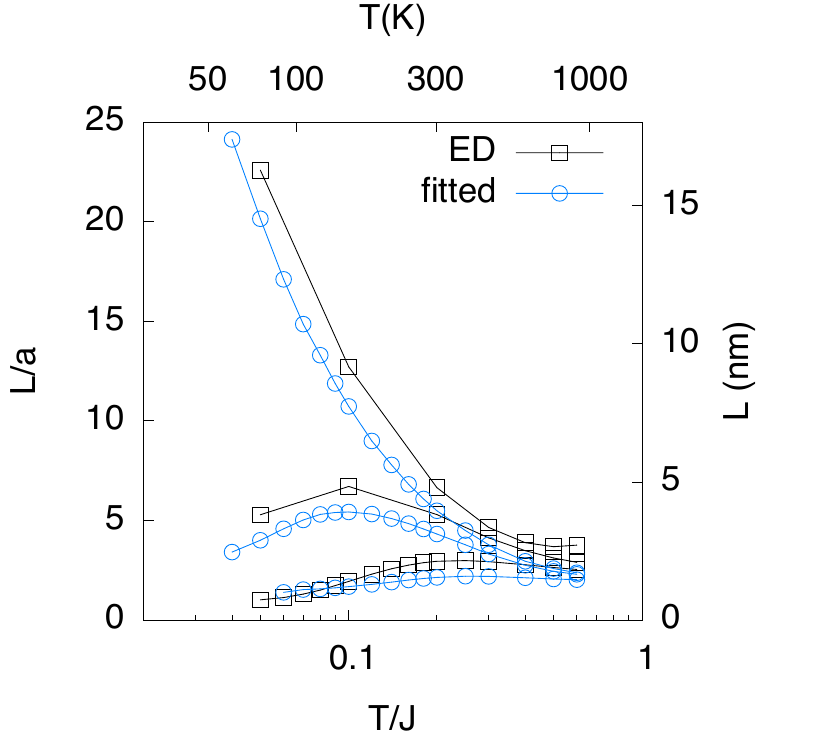}
\includegraphics[width=5.8cm]{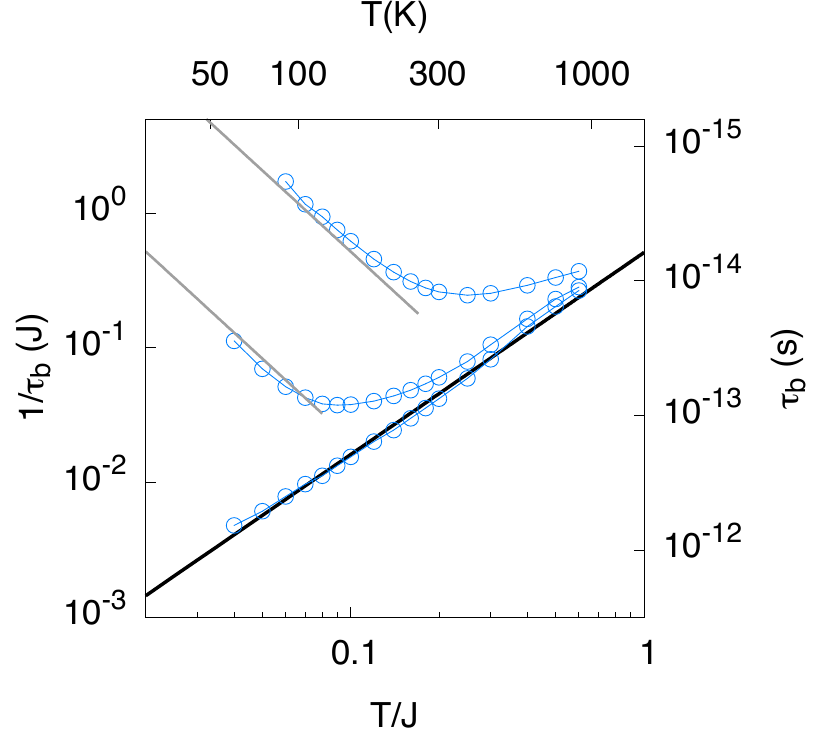}
\includegraphics[width=5.8cm]{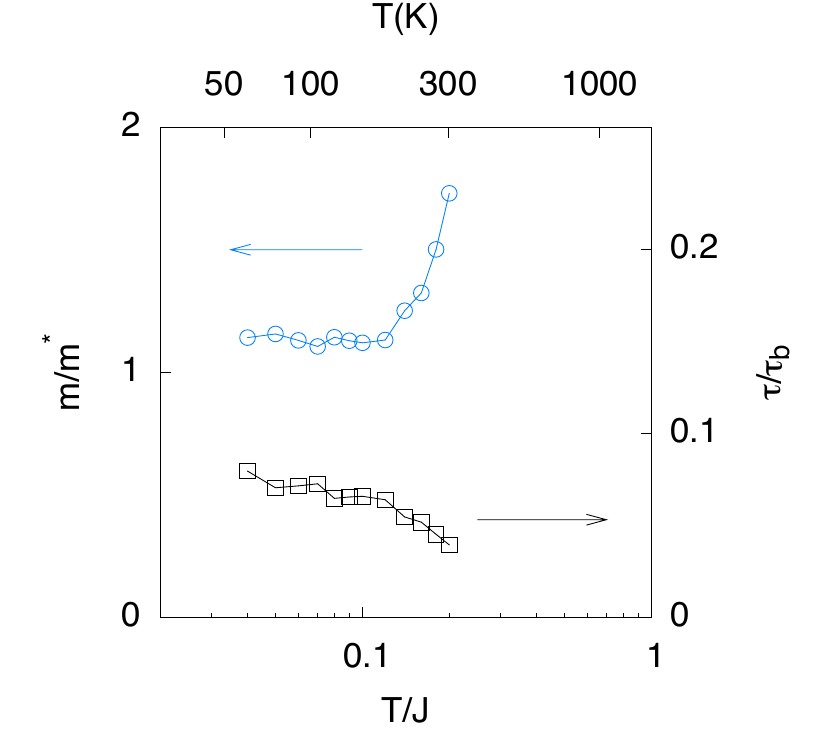}
   \caption{
   (a) The values of $L$ vs $T$ extracted from the fits (blue circles) are compared with 
   the exact values evaluated via ED (gray squares); 
   from top to bottom $\Delta=0;0.2;0.45 J$; the right axis units are obtained assuming $a=7.2\AA$ as appropriate for 
   rubrene;
  (b) Fitted values of the backscattering rate $1/\tau_b$ that controls electron localization
  (blue circles, from bottom to top $\Delta=0;0.2;0.45 J$); black triangles are estimates 
  obtained from the peak position via Eq. (\ref{eq:wstar2});
  the full line is a perturbative estimate valid for the intrinsic case $\Delta=0$ (see text). 
  The left axis is in units of $1/J$, the
  right axis in $s$, assuming $J=130meV$; (c) inverse effective mass estimated from Eq. (\ref{eq:L2mstar}) (left) 
  and ratio of the scattering timescales (clean limit $\Delta=0$ only).}
 \label{fig:fits}
 \end{figure*}
 \end{widetext}

 In Figs. \ref{fig:sigma}-a and \ref{fig:sigma}-b we show  
representative optical conductivity  spectra obtained from ED 
at different temperatures for intrinsic disorder
alone (a), and with added extrinsic disorder (b). The spectra are plotted in units of 
$\sigma_0=ne^2a^2/\hbar$.
In all cases, the optical absorption vanishes at $\omega\to 0$ as expected in a localized system
and shows a characteristic peak at a frequency that appears to be ruled by
the amount of disorder: the peak in (a) is progressively suppressed 
and moves to higher frequencies upon increasing the thermal
disorder. The peak is located at even higher frequencies in (b) due to the presence of 
additional extrinsic disorder [note the larger frequency interval in  panel (b)]. 
For sufficiently large disorder/temperature, 
a fully incoherent regime can be reached where the whole optical absorption lies below the 
the Mott-Ioffe-Regel value $\sigma_0$ (this value is $1$ 
in the units of Fig. \ref{fig:sigma}).

The full lines are fits to 
the phenomenological lineshape Eq. (\ref{eq:optcond}),
which satisfactorily reproduce the exact spectra  in the
relevant  region of the localization peak.  The  
localization length $L$ extracted from the fits is shown in Fig.  \ref{fig:fits}-a (blue circles).
The extracted values agree
quite well with the exact values given in Ref. \cite{Ciuchi12} (gray squares). It is important to stress that 
such exact values were obtained via a sum rule involving 
the whole optical conductivity spectrum, i.e. the optical response of the electrons
{\it at all frequencies} [see Eq. (10) in Ref. \cite{Ciuchi12}].
The quantitative agreement between the fitted and exact values implies that 
the present phenomenological fitting procedure
is able to carefully extract the localization length from 
the knowledge of the optical response
in the peak region alone. This is crucial  when 
it comes to the analysis of experimental data, where 
the spectrum at all frequencies is generally not known and the 
exact sum rule analysis for the determination of $L$ cannot be applied.

The fitted backscattering rate is shown in Fig.  \ref{fig:fits}-b (blue circles). 
In the intrinsic regime, 
the temperature dependence is consistent with $1/\tau_b\propto T^{3/2}$. 
From the results in Fig.  \ref{fig:fits}-b  we argue that in the present units the 
backscattering rate is quantitatively described by 
$1/\tau_b\simeq 3 \lambda T^{3/2}/J^{1/2}$, which is
shown as a full black line (we have checked that this result holds at different values of $\lambda$). 
This relation 
can be used in principle to determine experimentally the value of the electron-vibration coupling $\lambda$
from the value of $\tau_b$ fitted on the optical conductivity spectra
in sufficiently pure samples. 
A similar analysis yields $1/\tau_b\simeq (0.6 \Delta)^4 /(JT^2)$ in the extrinsic limit (strong 
extrinsic disorder, low temperatures, gray lines).
We note that 
use of the simple Eq. (\ref{eq:wstar2}) to extract the backscattering rate from the position of the peak alone 
gives results that are consistent with the fits, both 
in the intrinsic regime ($\Delta=0$) and for weak extrinsic disorder ($\Delta=0.2J$). 
Deviations arise for large extrinsic disorder and at low temperatures, i.e. where 
$ 0.3 \hbar/\tau_b \gtrsim k_BT $,
in which case Eq. (\ref{eq:wstar1}) should be used instead.

Eq. (\ref{eq:optcond}) also allows 
to extract the elastic scattering time $\tau$ from the optical conductivity data, provided that 
the overall disorder is not too large. Fits
 to the exact results from the microscopic one-dimensional model in the clean limit ($\Delta=0$)  
 yield $ 1/\tau \simeq 40 \lambda T^{3/2}/J^{1/2}$.
  Comparing this result to the value of $\tau_b$ given above, we 
  see that in the present one-dimensional model the elastic and backscattering timescales are related by an 
  approximately constant ratio $\tau/\tau_b\simeq 0.07$. This is shown in Fig. \ref{fig:fits}-c (right axis), and agrees
 with the Thouless argument given after 
  Eq. (\ref{eq:tauratio}).  We note that when the disorder becomes large, 
  the scattering rate $1/\tau$ becomes comparable with the bandwidth itself 
  and it is no longer possible to extract this parameter with sufficient 
  confidence within the present fitting procedure. This explains why the data of Fig. \ref{fig:fits}-c are limited to the
  clean case $\Delta=0$ and to temperatures $T<0.2J$.

 Finally, in Fig.  \ref{fig:fits}-c (left axis) we check the validity of Eq. (\ref{eq:L2mstar}) by plotting the 
 estimated band mass $m^*$ as obtained from the ratio  $L^2/[2k_BT\tau(\tau_b-\tau)]$, by substituting 
 the fitted values of $L,\tau$ and $\tau_b$ reported in  panels a and b. The estimated band mass 
 is expressed in units of the known band mass of the one-dimensional model, $m=1/2J$. For all temperatures
 $T\lesssim 0.15 J$ we have $m^*\simeq m$, while the departure observed at the highest temperatures 
can again be ascribed to an  inaccurate  fit of the elastic scattering rate when this becomes comparable to the bandwidth. 
 The agreement of the estimated $m^*$ with the known value means that Eq. (\ref{eq:L2mstar}) can be used 
 in practice to estimate the band mass when the localization length $L$ and the scattering timescales are known, or 
 alternatively to estimate $L$ from the known value of the band mass and the fitted values of the scattering timescales.

\section{Experimental analysis}

\subsection{Rubrene FETs}

\begin{figure}[h]
   \centering
\includegraphics[width=8.cm]{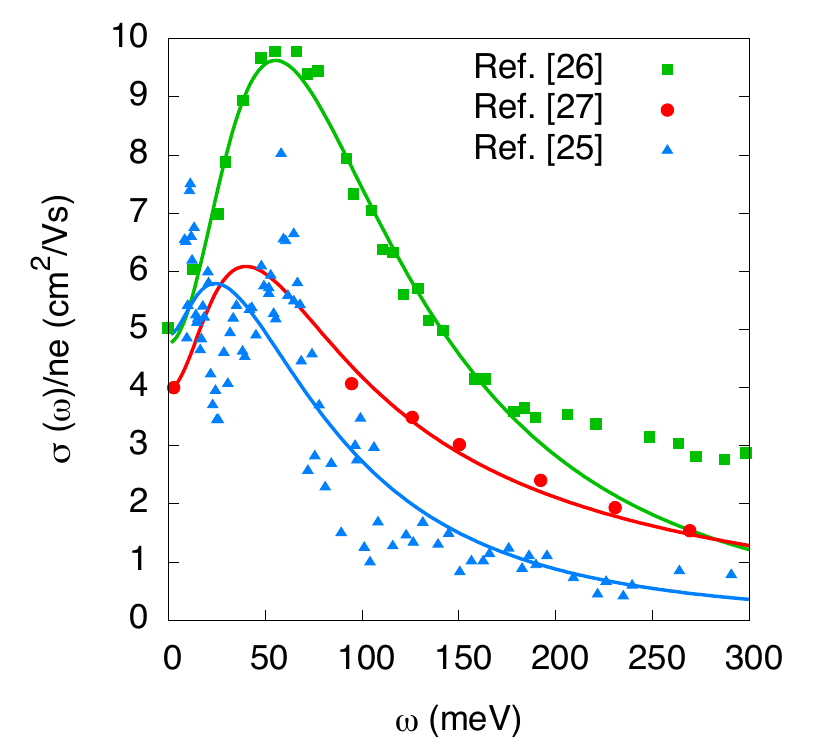}
   \caption{Available optical absorption data on rubrene FETs. The data are expressed in mobility units 
   and have been  rescaled to yield the FET mobility value in the zero frequency limit (see text).
    Full lines are fits of the data via Eq. (\ref{eq:optcond}).}
 \label{fig:exp}
 \end{figure}

Few optical absorption experiments on field-effect doped 
crystalline organic semiconductors have been reported in the literature\cite{Fischer,Li,Okamoto}, all of them 
performed on rubrene single crystals. Measurements of different groups, 
all taken at room temperature, are reproduced in Fig. \ref{fig:exp}.
The data have been rescaled here to 
%yield a common value $\mu=5cm^2/Vs$ in the zero frequency limit, 
recover, in the d.c. limit,  the mobility values that have been measured independently via the FET transfer characteristics 
(the reported FET mobility is $\mu=5cm^2/Vs$ in Refs.
\cite{Li,Fischer} and $\mu=4cm^2/Vs$ in Ref. \cite{Okamoto}). 
The different optical measurements exhibit considerable scatter, possibly related to 
differences in the experimental setups. However, all of them show (or are compatible with) an absorption
 peak around $\omega\simeq 50meV$, which can not be possibly reproduced within the semiclassical Drude model.
We argue that this feature is a fingerprint of the transient localization mechanism. 

We focus first on the data of Ref. \cite{Li}
measured in the direction of highest conduction and at the highest reported gate voltage. 
As illustrated in Fig. \ref{fig:exp}, the analytical formula 
Eq. (\ref{eq:optcond}) is able to 
closely reproduce the experimental absorption peak at $\omega\simeq 50meV$. 
The following parameters are obtained from the fitting procedure:  $1/\tau_{in}=13 meV$, $1/\tau_b=40 meV$, 
$1/\tau=195meV$.
The extracted inelastic scattering rate $1/\tau_{in}$ is 
consistent with the frequency of the intermolecular vibrations in rubrene,
$\omega_0=5-15 meV$,
as independently measured and theoretically calculated in Refs.  \cite{Ren,Girlando}. 
On the other hand, 
 since there are no independent measurements of the backscattering rate 
 available on the present device, 
we compare the extracted value $1/\tau_b=40 meV$ with the results of
the microscopic model calculation 
reported in Fig. \ref{fig:fits}-b. Taking $J=130$meV as representative for rubrene implies
$(1/\tau_b)/J=40/130=0.3$. 
It can be read from Fig. \ref{fig:fits}-b that this value at room temperature ($T/J=0.2$)   
corresponds to a degree of extrinsic disorder  $\Delta\simeq 0.45J=60$meV, which is
in the typical range observed in these devices. 
The extracted value of $\Delta=60$meV  implies that  carrier transport in the studied sample
is still far from the intrinsic regime.
This conclusion agrees with the fact that the  measured
FET mobility $\mu=5 cm^2/Vs$ is considerably lower than the highest values $>20 cm^2/Vs$
reported in rubrene-based FETs of higher purity \cite{Podzorov,Takeya}

We note that in Eq. (\ref{eq:optcond}) the global amplitude is proportional to the factor $nL^2$,  so that 
fits to the optical conductivity do not give access separately 
to the carrier density $n$ and the localization length $L$.
In clear, the localization length can only be obtained if $n$ is known independently from an independent measurement or, 
alternatively, if it can be estimated  from the known value of the band mass
(following the procedure described in Appendix A).
In the FET geometry used in Ref \cite{Li}, for example, the carrier density 
was determined from the known device capacitance  to $n=3.7\cdot 10^{12}$cm$^{-2}$, which allows us to 
determine $L/a=1.9$ from the fit in Fig. 5.  
This value is actually in very good agreement with the  value
predicted from the microscopic model at this temperature and this level of extrinsic 
disorder, $L/a=2.1$, as shown in Fig. \ref{fig:fits}-a.

 Tentative fits to the other available measurements can be attempted for qualitative purposes, 
even though it is more difficult to extract reliable quantitative parameters in these cases. 
To reduce the number of degrees of freedom in the fits we 
fix the value of the inelastic scattering rate to 
$1/\tau_{in}=13 meV$ as obtained previously, 
which is justified because the inter-molecular vibration frequencies
are not expected to vary from sample to sample.
The data of Ref. \cite{Fischer} exhibit considerable scatter, 
and the fit quality is not as good as in the previous case. A
 localization peak is well visible at $\omega\simeq 50$meV as in the data of Ref. \cite{Li}, 
but the large increase of conductivity at low frequency leads to a two-peak structure that 
cannot be well described by our formula  (the resulting peak position in the fit lies in between the
two maxima in the experimental data). 
Finally in Ref. \cite{Okamoto} only the high frequency tail of the absorption was measured, but 
not the region of the localization peak that is of interest to us. Still, our fit
with Eq. (\ref{eq:optcond}) does reveal the existence of a localization peak in the region where 
no data points are available, which provides an alternative interpretation to the one proposed
in Ref. \cite{Okamoto} based on the semiclassical Drude model.

\section{Concluding remarks}

We have derived a general 
phenomenological formula that describes the low-frequency optical absorption of
charge carriers in disordered systems, interpolating
between the Drude-like response of diffusive carriers and the finite-frequency 
peak shape expected in the presence of Anderson localization. 
Such Drude-Anderson  formula provides a useful alternative 
to the standard Drude model and to its known generalizations --- extended  Drude,  
Drude-Lorentz or Drude-Smith\cite{Smith} models  (see e.g. Ref. \cite{Ulbricht-RMP11}) ---
for the analysis of optical conductivity experiments. 
This analytical formula has been benchmarked
by comparing it with exact numerical results obtained on a microscopic model with on-site and inter-site
disorder that is believed to be relevant to high-mobility organic semiconductors, and has been shown to
give a physically transparent description of the phenomenon 
of transient localization.
We have then applied it to the analysis of the available experimental data in rubrene-based FETs, 
showing that these can be consistently and quantitatively interpreted within the transient localization scenario.

Interestingly, the same concept of transient localization that has been developed in recent years 
in the context of organic semiconductors
was also applied in the past to low-dimensional organic
{\it metals} such as the TCNQ salts \cite{Shante78,Gogolin75,Madhukar77,Gogolin82}.
Such  compounds have a  molecular structure very similar to the
organic semiconductors studied here, consisting of organic molecules weakly bound by 
van der Waals forces. 
Similar microscopic mechanisms are therefore expected to be at work, and it is 
not surprising that conductive properties 
comparable to those of crystalline organic semiconductors
are  commonly observed  in organic metals, at least in the high temperature 
range where electronic correlations are unimportant.
Indeed,  the d.c. conductivities in these materials typically exhibit a power law decrease with 
temperature, and  
room temperature  mobilities in the range  $\mu = 0.1-10cm^2/Vs$    
can be deduced from the reported conductivity values  at room temperature
(these are on the order of $\sigma = 10-1000 (\Omega cm)^{-1}$, see Ref. 
\cite{Graja} for a complete review of different compounds).

More recently, it has been independently suggested based on a combined analysis of 
the d.c. and optical conductivity
\cite{Gunnarsson} 
that the modest conductivity values observed \cite{Takenaka} in the two-dimensional organic superconductor 
$\theta$-ET$_2$I$_3$  
[$\sigma = 10 (\Omega cm)^{-1}$ at room temperature, again below the Mott-Ioffe-Regel limit] originate
from  the coupling of electrons to  molecular vibrations.
% rather than from electron-electron correlations. 
Our theory does provide support to this scenario, since both the lineshapes and temperature evolution 
of the optical absorption spectra of Ref. \cite{Takenaka} are comparable to those of  
our Fig. \ref{fig:sigma}-a.
The comparison can be made even more quantitative by fitting the  data 
reported in Ref. \cite{Takenaka}. For this we use the generalization of the optical conductivity formula Eq. (11) 
that applies to metallic systems in the degenerate limit, presented in Appendix C.
The extracted backscattering rate $1/\tau_b\simeq 14 meV$ ($\tau_b\simeq 3.1\cdot 10^{-13}s$) at room temperature
is right in range expected from electron-molecular vibration coupling (cf. Fig. \ref{fig:fits}-b). 
The similarities observed in the transport and optical properties 
of organic semiconductors, conductors and superconductors
 suggest that the transient localization scenario discussed here 
is a more general and common characteristic of the whole broad class of organic molecular crystals.

Finally, we mention that in a different class of low-dimensional materials --- carbon nanotubes ---
an ubiquitous optical absorption peak has been reported by several groups  in the far infrared range
\cite{Kampfrath,Ichida,Ulbricht-RMP11}, whose microscopic origin could possibly be related to the localization phenomena 
described in this work.

\appendix

\section{High frequency limit: semiclassical dynamics}
The  semiclassical diffusion of electrons corresponds
to the assumption that velocity correlations
%, that would be present in the absence of scattering, 
are destroyed after a relaxation time $\tau$.
This assumption is appropriate when the particles
are only weakly scattered by disorder. 
To illustrate this we take the simple model:
\begin{eqnarray}
& & C(t)=C(0)e^{-t/\tau}  \label{eq:Drude}
\\
& & \Delta X^2(t)=  C(0)\tau 
\left\lbrack t-\tau +\tau (1-e^{-t/\tau})
\right\rbrack. \label{eq:X2Drude}
\end{eqnarray} 
Eq. (\ref{eq:Drude}) assumes that correlations in the absence of scattering   
are  independent of time, which is certainly valid for classical trajectories.
The particle spread Eq. (\ref{eq:X2Drude}) has been obtained by double 
integration of the correlation function over time. Both quantities are
illustrated in Fig. \ref{fig:XandC} as dashed lines.
Eq. (\ref{eq:X2Drude}) describes an initial ballistic behavior 
$\Delta X^2(t)=C(0) t^2/2$ followed by diffusion $\Delta X^2(t)= C(0)\tau t$ (the corresponding
diffusivity is $D=C(0)\tau/2$). Performing the Fourier transform yields
$C(\omega)= C(0)/(1/\tau-i\omega)$ and the corresponding optical absorption is obtained via 
Eq. (\ref{eq:sigmatransf}). We note that this derivation recovers the usual Drude response 
in the high temperature limit: in this case we have 
$C(0)=2\langle V^2 \rangle = 2N k_BT/m^*$ from the equipartition principle, and  
expanding Eq. (\ref{eq:sigmatransf}) for $k_BT\gg \hbar\omega$  we have 
\begin{equation}
\label{eq:Drude}
\sigma(\omega)=(ne^2\tau/m^*)Re \left[ \frac{1}{1-i\omega \tau}\right].
\end{equation}
The Drude expression Eq. (\ref{eq:Drude}) correctly describes the optical response of the complete model Eq. 
(\ref{eq:optcond}) in the high frequency limit where $\omega\gg 1/\tau_{in},1/\tau_b$, as shown in Fig. 2.
In this case the second term between brackets in Eq. (10) is negligible and the first term is
precisely of the Drude form with the correct prefactor.  
This can be easily demonstrated by observing that $\tanh(\hbar\omega/2k_BT)/\hbar\omega=1/2k_BT$ at high temperatures 
and using the relation
$L^2= D_{sc}\tau_b$ [see Eq. (\ref{eq:tauratio})] with $D_{sc}=2\langle V^2 \rangle \tau = 2k_BT\tau/m^*$, which yields 
Eq.  (\ref{eq:L2mstar}).

More generally, the knowledge of the semiclassical diffusivity 
is extremely useful when it comes to analyzing experimental data where the absolute number of carriers is
not known, because it provides a prescription to to estimate  the carrier number provided that the
effective mass $m^*$ is known. This is a non-trivial issue, as fits of the optical conductivity via 
Eq. (\ref{eq:optcond}), which is proportional to the product $nL^2$,  cannot in principle extract $n$ and $L$ separately.
In practice, to extract $n$ one replaces $L^2$ in  Eq. (\ref{eq:optcond}) by its expression
Eq. (\ref{eq:L2mstar}), so that the prefactor of the optical conductivity now only depends on $n$ via the 
known values of $m^*$ and $\tau$.
  
A generalization of Eq. (\ref{eq:Drude})
that recovers the usual Drude formula in the degenerate limit at $T=0$ is reported in Appendix C and 
can be found in Refs. \cite{Mayou00,Gruner}.

\section{Quantum diffusion in the phenomenological model}
We provide here the full expression for the quantum diffusion in the phenomenological model
Eq. (\ref{eq:Cdyn}):
\begin{eqnarray}
& & \Delta X^2_{RTA}(t)=2NDt+ \label{eq:X2dyn}\\
& & + \frac{NL^2}{\tau_b-\tau} \left\lbrack
\frac{(1-e^{-\tilde\gamma_b t})}{\tau_b\tilde\gamma_b^2}-
\frac{(1-e^{-\tilde\gamma t})}{\tau\tilde\gamma^2}
\right\rbrack
\nonumber 
\end{eqnarray} 
where we have defined the quantities $\tilde\gamma=(1/\tau+1/\tau_{in})$ and
$\tilde\gamma_b=(1/\tau_b+1/\tau_{in})$, and
\begin{equation}
\label{eq:Dtransloc}
D=\frac{1}{(1+\tau/\tau_{in})(1+\tau_b/\tau_{in})}\frac{L^2}{2\tau_{in}} 
\end{equation} 
is the diffusion constant at long times. Eq. (\ref{eq:Ddyn}) of the main text is obtained by
taking the limit of slow disorder fluctuations, $\tau_{in}\gg \tau_b,\tau$.

The above expression allows us to define the {\it transient} localization length as
\begin{equation}
L^2(\tau_{in})=\frac{L^2}{(1+\tau/\tau_{in})(1+\tau_b/\tau_{in})}
\label{eq:transloc}
\end{equation} 
so that $D=L^2(\tau_{in})/2\tau_{in}$.
The quantity in Eq. (\ref{eq:transloc})  
is the spread of the localized electron wavefunction 
on a timescale $\tau_{in}<\infty$, as defined in Eq. (7) of Ref. \cite{Ciuchi12}. 
It is clear from the above definition that $L(\tau_{in})<L$ for all
finite $\tau_{in}$.
%It is this quantity, not $L$,  that 
%enters the formula Eq. (\ref{eq:Ddyn}) for the diffusivity and Eq. (\ref{eq:mu}) 
%for the mobility. 
In practical cases when the timescales $\tau_{in},\tau_b$ and $\tau$ are
not well separated, this can make a sizable correction to the mobility and 
the full expression Eq. (\ref{eq:transloc}) 
should be used in Eqs. (\ref{eq:Ddyn}) and (\ref{eq:mu}) instead of the bare $L$.

\section{Case of a degenerate electron system}

As explained in the main text the  real part of the optical conductivity is obtained as 
\begin{equation}
\sigma(\omega)=\frac{e^2\tanh(\frac{\hbar\omega}{2k_BT})}{\hbar \omega \Omega} Re C_+(\omega), 
\label{eq:sigmatransfapp}
\end{equation}
where $e$ is the electron charge, $\Omega$ is the system volume and 
 $C_+(\omega)=\int_0^\infty e^{i\omega t} C_+(t) dt$. $C_+(t)$ is the time-dependent velocity correlation function for the  
 N-electron  system.
When the electron system  is non degenerate and follows the 
Maxwell-Boltzmann distribution the velocity correlation $C_+$ has a simple expression in terms of 
the velocity correlation of a single particle, being  N-times the thermodynamical 
average of the velocity correlation function of one electron alone. The approach developed in the main text  
is based on the assumption that the velocity correlation for one electron can be well represented by 
Eqs. (\ref{eq:Cloc}) and (\ref{eq:X2loc}).

In the case of a degenerate electron system the many-body velocity correlation 
function is no longer equal to N times the thermodynamical average of
the correlation function of one electron alone,  due
to quantum statistics. Yet at low temperature and low frequency  $k_BT\ll \hbar
\omega \ll E_F$, 
where $E_F$ is the Fermi energy, 
it is
possible to express the conductivity in terms of the velocity correlation
function $C_+(E_F,t)$ for states at $E_F$. The derivation 
can be found  in Refs. \cite{Mayou00,Gruner}. 
At frequencies much smaller than the Fermi energy the conductivity is then given  by:
\begin{equation}
\sigma(E_F,\omega)=e^2N(E_F) \ Re \int_0^\infty e^{i\omega t}C_+(E_F,t)dt
\label{eqtransloc}
\end{equation}

The above equation is similar to Eq. (\ref{eq:sigmatransfapp}) except that 
$C_+(t)$ is replaced by $C_+(E_F,t)$, and the prefactor now 
involves  $N(E_F)$, the density of states per unit volume
and spin at the Fermi energy (a perfect spin degeneracy is assumed here). For
the velocity correlation at the Fermi energy we assume that the behavior is
analogous  to that given in
Eq. (\ref{eq:Cloc}) in the main text, i.e.: 
\begin{eqnarray}
C(E_F,t)&=& \frac{C(E_F,0)}{1/\tau-1/\tau_b}\left\lbrack 
\frac{1}{\tau}e^{-t/\tau}-\frac{1}{\tau_b}e^{-t/\tau_b}\right\rbrack 
\label{eq:Clocapp}
\end{eqnarray}
\begin{eqnarray}
 \Delta X^2(E_F,t)&=& \frac{C(E_F,0)}{1/\tau-1/\tau_b}
 \label{eq:X2locapp}\\
& & \times \left\lbrack 
\tau_b(1-e^{-t/\tau_b})-\tau(1-e^{-t/\tau})
\right\rbrack \nonumber
\end{eqnarray}
The velocity correlation at zero time $C(E_F,0)$ is typically given by $2
V_{F}^2 /d$  where $V_F$ is the Fermi velocity and $d$ the dimensionality. The
above Eq. (\ref{eq:X2locapp}) describes the localization of electrons at the Fermi
energy in a way similar to that described in the main text for non-degenerate electrons. 

The effect of inelastic scattering can be introduced in the spirit of the RTA as in 
Eq. (8) of the main text:
\begin{equation}
C_{RTA}(E_F,t)=C_+(E_F,t)e^{-t/\tau_{in}} \label{eq:Cin}.
\end{equation} 
This leads to the following expression for the complex conductivity:
\begin{eqnarray}
\sigma(E_F,\omega)&=&e^2N(E_F)\frac{C(E_F,0)}{1/\tau-1/\tau_b}
\label{eqtransloc2}\\
& & \times  \left\lbrack
\frac{1}{1+\tau/\tau_{in}-i\omega\tau}-\frac{1}{1+\tau_b/\tau_{in}-i\omega\tau_b}
\right\rbrack \nonumber 
\end{eqnarray}
Except for the different prefactor, this expression is analogous to 
Eq. (\ref {eq:optcond})  given in the main
text  and can therefore describe  
a finite frequency localization peak. When the inelastic
scattering time $\tau_{in}$ tends to infinity then the zero frequency
conductivity vanishes. When the inelastic scattering time is finite the zero
frequency conductivity is finite and varies proportionally to $1/\tau_{in}$ in
agreement with the Thouless regime, as discussed in the main text.
The usual Drude formula is recovered in the limit $\tau_b\to \infty$.

\begin{acknowledgements}
We thank Prof. R. Martel for stimulating discussions on the infrared response of carbon nanotubes.
\end{acknowledgements}


\begin{thebibliography}{99}
\bibitem{note-geometrical}
Electron localization, embodied in the condition $\mathcal{D}(t\to\infty)=0$, 
has a direct geometrical interpretation that can be read directly from Fig. \ref{fig:XandC}-b: 
because the long-time diffusivity is proportional to the integral of the velocity correlation function, 
in a localized system the positive area under the curve $C(t)$ is exactly compensated by the 
negative backscattering part. This cancellation is destroyed by the exponential decay in Eq. (\ref{eq:Cdyn}),
restoring a finite diffusivity at long times.
\bibitem{C0} The value of the initial constant is also fixed by the model to  $C(0)=NL^2/(\tau_b\tau)$.
We point out here that 
this value does not correspond to the actual short-time limit
of the correlation function, $C(0)=\langle \hat{V}^2 \rangle$. 
This discrepancy arises from the fact that the simple model Eq. (\ref{eq:Cloc}) is devised to provide a 
good description of the carrier dynamics at frequencies 
relevant for  diffusion and localization (intermediate and long times), but it has no information 
on the high frequency behavior related to the exact form of the electronic bands (short times).
Strictly speaking, Eq. (\ref{eq:Cloc})  holds at times $t> t^*$ with 
$t^*$ a time of the order of the intermolecular transfer time $\hbar/J$.
\end{thebibliography}
\end{document}